\documentclass{article}
\usepackage{spconf,amsmath,graphicx,bm,amssymb}

\usepackage{xcolor}

\title{Deep Learning for Frame Error Probability Prediction \\ in BICM-OFDM Systems}
\name{Vidit Saxena\textsuperscript{1,2}, Joakim Jald\'{e}n\textsuperscript{1}, Mats Bengtsson\textsuperscript{1},
 and Hugo Tullberg\textsuperscript{2}
\thanks{This work was partially supported by the Wallenberg Autonomous Systems and Software Program (WASP).}}
\address{\textsuperscript{1}Department of Information Science and Engineering, KTH, Stockholm \\
\textsuperscript{2}Ericsson Research, Stockholm}
\begin{document}
\ninept
\maketitle
\begin{abstract}
In the context of wireless communications, we propose a deep learning approach to learn the mapping 
from the instantaneous state of a frequency selective fading channel to the corresponding frame error probability (FEP)
for an arbitrary set of transmission parameters.
We propose an abstract model of a bit interleaved coded modulation (BICM) orthogonal frequency division multiplexing (OFDM)
link chain and show that the maximum likelihood (ML) estimator of the model parameters estimates the true FEP distribution.
Further, we exploit deep neural networks as a general purpose tool to implement our model 
and propose a training scheme for which, even while training with the binary frame error events (i.e., ACKs / NACKs), 
the network outputs converge to the FEP conditioned on the input channel state.
We provide simulation results that demonstrate gains in the FEP prediction accuracy with our approach as compared to the
traditional effective exponential SIR metric (EESM) approach for a range of channel code rates,
and show that these gains can be exploited to increase the link throughput.

\end{abstract}
\section{Introduction}
\label{sec:intro}

The efficiency of a radio link depends on its ability to adapt to the stochastic radio channel conditions that typically vary
over time (i.e., fading) as well as over the signal bandwidth (i.e., frequency selectivity.
Practical radio systems solve this problem by selecting the transmission parameters in each frame to fulfil
some optimality criteria such as desired throughput, latency, etc.~\cite{B_Wireless_Molisch}.
In this paper we investigate the problem of frame error probability (FEP) prediction for bit-interleaved coded modulation 
orthogonal frequency division multiplexing (BICM-OFDM) systems~\cite{J_Caire_BICM}~\cite{C_Agrawal_OFDM}, which is an essential step in the selection
of optimal tranmission parameters.
Owing to their flexibility and performance, BICM-OFDM systems have been widely adopted by most of the modern radio air interfaces
including those for local wireless area networks (e.g., WiFi) and for cellular communication such as Long Term Evolution for 4G, and more recently, New Radio for 5G~\cite{J_Ali_NR}.

In general for frequency selective channels, it is intractable to compute the FEP conditioned on the frame channel state characterized by the
received per-subcarrier signal to interference and noise ratios (SINRs).
Therefore, several approximate teachniques for FEP prediction have been developed that compress the channel state vector to
an approximate effective channel state scalar, which is mapped to pre-computed FEP values 
stored within lookup tables~\cite{J_Nanda_Eff_SNR}~\cite{C_Blankenship_Link_Error}~\cite{M_EESM}.
In addition to the loss of channel state information in this approach, the choice of compression function is 
somewhat arbitrary and the optimality of the predicted FEP has only been studied empirically~\cite{C_Astely_LA}.

As an alternative to the effective SINR approach, supervised learning techniques that directly learn the mapping between per-subcarrier SINRs and the 
corresponding FEP have been developed recently~\cite{J_Daniels_2011}~\cite{C_Yun_SVM_LA}. 
In the traning phase of these techniques, the per-subcarrier SINRs for several channel realizations 
along with their Monte Carlo simulated FEPs are used to iteratively train the model parameters until some convergence criteria is satisfied.
With sufficient training, these techniques have been shown to improve the realized link throughput in BICM-OFDM systems
compared to the effective SINR approach. However, these techniques do not provide any insight into the optimality of the
trained models.

In this paper, we cast the FEP prediction problem as a probabilistic binary classification task, 
where the classes correspond to frame error and success events (i.e., NACKs and ACKs) respectively, and make the following three main contributions: 
(i) We propose an abstract model of the BICM-OFDM link chain where the observations are the 
frame channel states and their binary frame error events and show that, in the limit of infinite training samples, 
the maximum likelihood (ML) estimator of the model parameters estimates the true FEP distribution,
(ii) we use this model to develop a supervised learning approach for FEP prediction based on deep neural networks,
where the training phase requires only the observed channel states and binary frame error events, 
thus eliminating the need for tedious Monte Carlo simulations to measure the FEP,
and (iii) we provide numerical results to show that our approach improves the FEP prediction accuracy 
compared to one widely studied effective SINR approach, and exploit these predictions to improve the realized link throughput.

The rest of this paper is organized as follows: In Sec.~\ref{sec:sys-model-ml}, we describe
a typical BICM-OFDM wireless communication link and propose an abstract system model along with an ML estimator of the model parameters.
Next in Sec.~\ref{sec:fep-pred}, we summarize a commonly studied effective SINR approach for FEP prediction and introduce our
deep learning approach. Finally in Sec.~\ref{sec:Evaluations}
we present simulation results for FEP prediction accuracy and realized throughput for the two considered
approaches and in Sec.~\ref{sec:Conclusion}, we conclude the paper.

\section{BICM-OFDM System and ML Estimation}
\label{sec:sys-model-ml}

\begin{figure*}
\centering
\includegraphics[width=\textwidth]{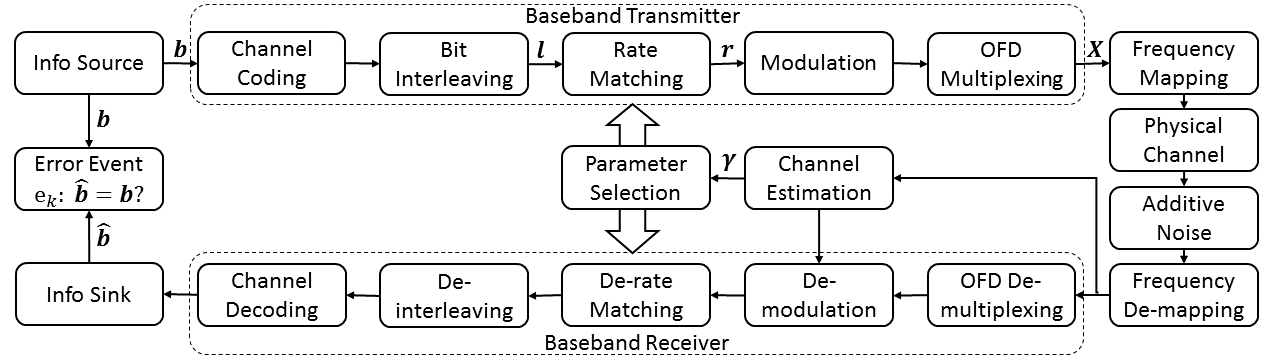}
\caption{Block diagram of the BICM-OFDM link chain considered in this paper. The parameter selection module exploits knowledge of the 
instantaneous channel state to select one out of several possible transmission parameter configurations in each frame.}
\label{fig:system-model}
\end{figure*}

\subsection{System Model}
\label{sec:SystemModel}

We consider a BICM-OFDM link chain similar to the LTE downlink and illustrate its block diagram in Fig.~\ref{fig:system-model}.
Here, a ``transport block", $\bm{b}=(b_1,\dots,b_T)$ of information bits is first encoded by a channel encoder and 
subsequently bit-interleaved by a random interleaver to generate the bit sequence $\bm{l}=(l_1,\dots,l_L)$. 
The interleaved bits are then used to generate the ``rate-matched" bit sequence $\bm{r}=(r_1,\dots,r_{MSJ})$ according to
$r_i = l_{i\text{ mod } L}, i=1,\dots,MSJ$,
where $M$ is the number of transmission subcarriers, $S$ is the number of frame OFDM symbols, and $J$ is the modulation order.
The channel code rate is therefore $\mathcal{R}=T/MSJ$. 
The rate matched bits are mapped onto $MS$ modulated symbols by a labeling function that assigns one out of $2^J$  
complex-valued constellation symbols to each $J$-tuple of bits. Finally, a length-$M$ IFFT operation
is applied on each group of $M$ modulated symbols to generate the frame OFDM symbols $\bm{X}=(\bm{x}_1,\dots,\bm{x}_S)$ and mapped
onto physical resources.

The frame OFDM symbols are transmitted over the physical channel resulting in the received signal
$\bm{y}_s = \bm{h}\odot\bm{x}_s + \bm{g}_s,s=1,\dots,S,$
where $\odot$ denotes the Hadamard product, $\bm{h}$ is the complex-valued vector of channel coefficients in the 
frequency domain, and $\bm{g}_s\sim \mathcal{N}^{M\times 1}(0,\sigma^2)$ is  i.i.d. noise.
We assume that the channel vector remains constant for the frame OFDM symbols (i.e., the channel is block fading). 
At the receiver, each received OFDM symbol $\bm{y}_s$ is multiplied by the elementwise inverse of the estimated frequency-domain channel vector,
followed by a length-M FFT operation for OFDM de-multiplexing. The de-multiplexed symbols are then mapped onto 
soft values through an inverse labeling operation, de-interleaved, and decoded by the channel decoder to generate the 
reconstructed bit sequence $\hat{\bm{b}}=(\hat{b}_1,\dots,\hat{b}_T)$. 
We define the binary frame error event at the receiver as
\begin{align}\label{eq:error-event}
e = \left\{
		\begin{array}{ll}
			0  & \text{if }\quad \hat{\bm{b}} =  \bm{b}\\
			1  & \text{if }\quad \hat{\bm{b}} \neq \bm{b}
		\end{array}.
	\right.
\end{align}

\subsection{ML Estimation }
\label{sec:ML-estim}

\begin{figure}[ht]
\centering
\includegraphics[width=\columnwidth]{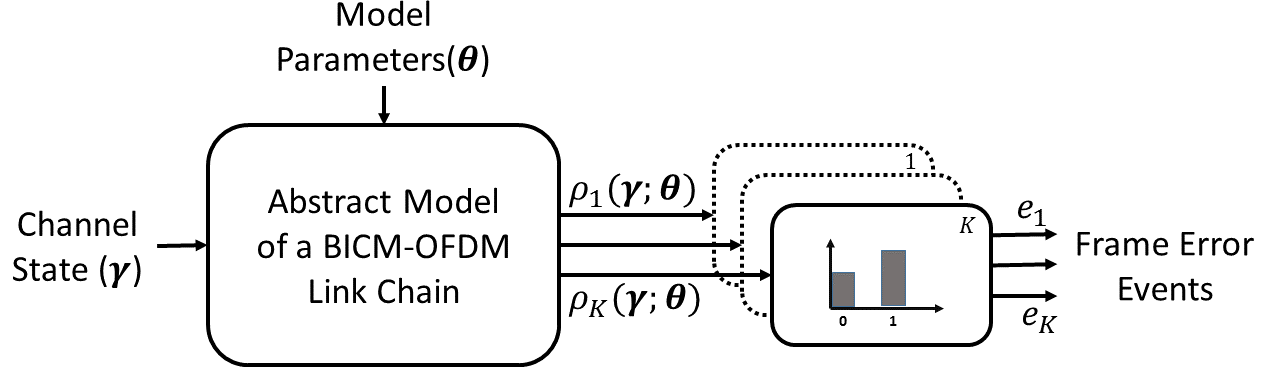}
\caption{Abtract model of a BICM-OFDM link chain that maps the observed channel state to the frame error events for $k=1,\dots,K$ transmission parameter configurations.}
\label{fig:simplified-sys-model}
\end{figure}

The BICM-OFDM link chain described above can be approximated as a stochastic non-linear function that generates a frame error event with an unknown probability
distribution for the frame channel state and a particular choice of transmission parameters.
In Fig.~\ref{fig:simplified-sys-model}, we illustrate an abstract model of the BICM-OFDM link chain,
which is parameterized by the model parameters $\bm\theta$ and maps the observed channel state characterized by the received per-subcarrier channel SINRs, 
$\bm\gamma=(\gamma_1\dots,\gamma_M)$, to the observed frame error event, .i.e.,
\begin{align}
\label{eq:bern-dist}
P_{E_k|\bm\Gamma}(e_k|\bm{\gamma};\bm{\theta}) = \rho_k^{e_k} (1-\rho_k)^{1-e_k},
\end{align}
where the $k\in 1,\dots,K$ denotes the $k^\text{th}$ transmission parameter configuration. 
Here, $\rho_k=\rho_k(\bm\gamma;\bm\theta)=P_{E_k|\bm\Gamma}(E_k=1|\bm\gamma;\bm\theta)$ is the conditional frame error probability (FEP) .
In the rest of this section, we show that the ML estimator of the model parameters asymptotically 
estimates the true conditional FEPs.

The ML estimator~\cite{B_EST_THEORY} of the model parameters for $n=1,\dots,N$ frame realizations is defined as
\begin{align}\label{eq:ml-estimator}
\hat{\bm{\theta}}^\text{ML} &= \arg\max_{\hat{\bm{\theta}}}\sum_{k=1}^K\prod_{n=1}^NP_{E_k,\Gamma}(e_k^n,\bm{\gamma}^n;\hat{\bm{\theta}}) \nonumber \\
                     			     &= \arg\max_{\hat{\bm{\theta}}}\int\sum_{k=1}^K\sum_{n=1}^N\ln P_{E_k|\Gamma}(e_k^n|\bm{\gamma}^n;\hat{\bm{\theta}})P_{\bm\Gamma}(\bm\gamma)d\bm\gamma \nonumber \\
                                                &\triangleq \arg\max_{\hat{\bm{\theta}}}\mathcal{C}(\hat{\bm{\theta}}), \text{where } \\
\mathcal{C}(\hat{\bm{\theta}}) &= \sum_{k=1}^K\left(\frac{1}{N}\sum_{n=1}^N\ln P_{E_k|\Gamma}(e_k^n|\bm{\gamma}^n;\hat{\bm{\theta}})\right) \label{eq:cost-func}
\end{align}
is the cost function to be maximized, and we have used the fact that the channel state is independent 
of the model. In the limit of infinite training samples, it follows by the law of large numbers that
\begin{equation}\label{eq:ML-KL-diverg}
\mathcal{C}(\hat{\bm{\theta}}) \xrightarrow{N\to\infty} \sum_{k=1}^KE\{\ln P(E_k|\bm{\Gamma};\hat{\bm{\theta}})\} \,
\end{equation}
where $P(E_k|\bm{\Gamma};\hat{\bm{\theta}}) \triangleq P_{E_k|\bm{\Gamma}}(E_k|\bm\Gamma;\hat{\bm{\theta}})$ for brevity, and 
where the expectation is taken over $P(E_k,\bm{\Gamma};\bm{\theta })$. We now subtract and add the true pmf to the r.h.s. of Eq.~(\ref{eq:ML-KL-diverg}) to obtain
\begin{align}\label{eq:ML-est}
\mathcal{C}(\hat{\bm{\theta}})  &= \sum_{k=1}^KE\{\ln P(E_k|\bm{\Gamma};\hat{\bm{\theta}}) - \ln P(E_k|\bm{\Gamma};\bm{\theta}) + \ln P(E_k|\bm{\Gamma};\bm{\theta})\} \nonumber \\
					    &= -\sum_{k=1}^KE\left\{\ln\frac{P(E_k|\bm{\Gamma};\bm{\theta})}{P(E_k|\bm{\Gamma};\hat{\bm{\theta}})}\right\} +  \sum_{k=1}^KE\left\{\ln P(E_k|\bm{\Gamma};\bm{\theta})\right\}.
\end{align}

The second term in (\ref{eq:ML-est}) is independent of the argument to be maximized. Further we observe that by multiplying and dividing the first term with 
probability distribution $P_{\bm\Gamma}(\bm\gamma)$, we obtain
\begin{align}\label{eq:ML-KL-div}
& \sum_{k=1}^KE\left\{\ln\frac{P(E_k|\bm{\Gamma};\bm{\theta})P(\bm\Gamma)}{P(E_k|\bm{\Gamma};\hat{\bm{\theta}})P(\bm\Gamma)}\right\}
		= \sum_{k=1}^KE\left\{\ln\frac{P(E_k,\bm{\Gamma};\bm{\theta})}{P(E_k,\bm{\Gamma};\hat{\bm{\theta}})}\right\} = \nonumber \\
	           & \sum_{k=1}^K\text{KL}\left\{P(E_k|\bm{\Gamma};\bm{\theta})||P(E_k|\bm{\Gamma};\hat{\bm{\theta}})\right\},
\end{align} 
where $\text{KL}(\cdot||\cdot)$ is the Kullback-Liebler divergence (KLD) between the true and estimated pmfs. Given that the KLD is non-negative, and equal to zero if and only if $P(E_k|\bm{\Gamma};\bm{\theta}) = P(E_k|\bm{\Gamma};\hat{\bm{\theta}})$, it follows that the ML estimator converges to the true FEP distribution in the limit of large $N$. Note however that $P(E_k|\bm{\Gamma};\bm{\theta}) = P(E_k|\bm{\Gamma};\hat{\bm{\theta}})$ does not necessarily imply that $\hat{\bm{\theta}}^\text{ML} = \bm{\theta}$ as the ML estimate of $\bm{\theta}$ may be non-unique.

\section{FEP Prediction Techniques}
\label{sec:fep-pred}

\subsection{Effective SINR Approach}
\label{subsec:CompressFn}

In this subsection we outline the widely studied exponential effective SIR metric (EESM) approach, where the 
the channel state characterized by the per-subcarrier SINRs is compressed to a scalar ``effective" 
SINR for an equivalent AWGN channel~\cite{C_Astely_LA}. 
The FEP for the $k^{th}$ transmission parameter configuration is then predicted to be
\begin{align}
\label{eq:eff-awgn-sinr}
\hat{\rho}_k^\text{ESM}(\bm\gamma) &= \rho_k^\text{AWGN}(g_k(\bm\gamma)),\\
\text{where }g_k(\bm\gamma) &=\beta_k\log\left(\frac{1}{M}\sum_{m=1}^M \exp\left(\frac{\gamma_m}{\beta_k}\right)\right)
\end{align}
is the EESM for channel state $\bm\gamma$, and $\beta_k$ is a tunable parameter.
For the frequency selective channels commonly observed in practical systems, $g_k(\bm\gamma)$ amounts to a
lossy compression of the channel state vector, since the original channel state can no longer be recovered.
The FEP for the equivalent AWGN channel, $\rho_k^\text{AWGN}(g_k(\bm\gamma))$, is obtained 
by interpolating between several Monte Carlo simulated FEP values for the AWGN channel.
The optimal values of the tuning parameters $\beta_k$ can be determined by minmizing the Euclidean distance 
between the predicted FEP and observed frame errors for $n=1,\dots,N$ training frames, i.e., 
\begin{align}
\beta^\text{opt}_k&=\arg\min_{\beta_k} \sum_{n=1}^N|\hat\rho_k^\text{n, ESM}-e_k^n|^2.
\end{align}
As $N$ grows to infinity, $\beta_k^\text{opt}$ also minimizes the least squares error between the estimated and true FEPs, however,
we skip the proof here owing to space constraints.

\subsection{Deep Learning Approach}
\label{subsec:NeuralNet}

The EESM approach described earlier relies on a scalar approximation of the channel state, which is obtained through a lossy compression 
and therefore does not guarantee optimality of the the corresponding FEP prediction.
In this subsection we describe our approach for FEP prediction based on deep neural networks, which discriminatively learns the mapping between the 
(uncompressed) channel state vector and the corresponding FEPs for multiple transmission parameter configurations.

\begin{figure}[ht]
\centering
\includegraphics[width=\columnwidth]{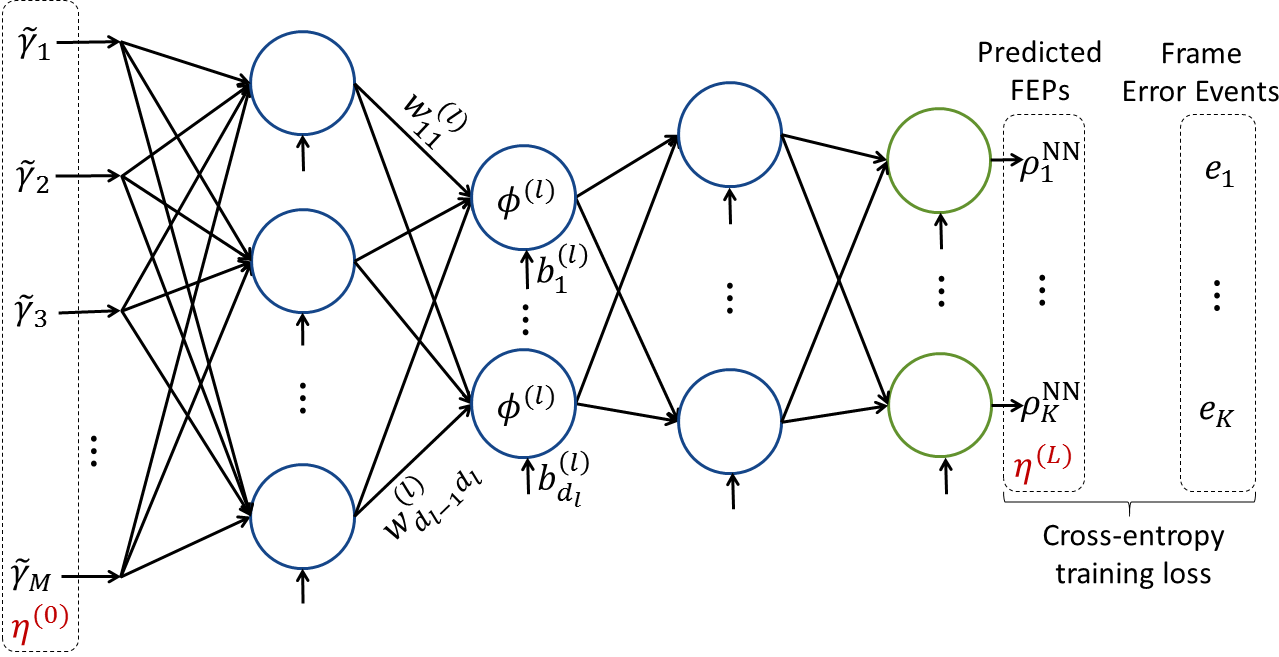}
\caption{Neural network layout that maps the channel state characterized by per-subcarrier
SINR values to the FEP for each of the $K$ transmission parameter configurations.}
\label{fig:NNLayout}
\end{figure}

Neural networks have long been known as a powerful tool for approximating a wide range of highly non-linear functions,
however, their acceptance for implementation in practical systems has been limited by an insufficient understanding of the models that they learn from training data.
Although a complete understanding of neural networks is still a topic of active research, several recent breakthroughs related to deep neural networks 
coupled with cheap computational power have led to drastic performance improvements for several challenging problems~\cite{B_DEEPLEARNING}. 
In this paper, we consider the fully connected $L-$layered deep neural network illustrated in Fig.~\ref{fig:NNLayout}, 
for which the output of the $l^{th}$ ``hidden layer" with dimension $d_l$ can be described as
\begin{align}
\bm{\eta}^{(l)}&= \bm{\phi}^{(l)}\left(\bm{W}^{(l)}\bm{\eta}^{(l-1)}+\bm{b}^{(l)}\right), 
\end{align}
where $\bm{W}^{(l)}$ is the trainable $d_{l-1}\times d_l$ weight matrix,
$\bm{b}^{(l)}$ is the trainable $d_l\times 1$ bias vector, and $\bm{\phi}^{(l)}$ is a fixed non-linear ``activation" function. By simply substituting the system parameters $\bm{\theta}$ with neural network weights and biases, we allow the neural network to learn the set of mappings $\rho_k(\bm{\gamma};\hat{\bm{\theta}})$ for $k=1,\ldots,K$ from data.

It has been shown previously that an ordering of the per-subcarrier SINRs
can be used to sufficiently parameterize the frame error rate while reducing the training requirements~\cite{J_Daniels_2011}.
Therefore in this paper, we use the sorted per-subcarrier SINR vector, $\widetilde{\bm\gamma}$, 
as the input to the network, i.e., $\bm{\eta}^{(0)} \triangleq \widetilde{\bm\gamma}=(\widetilde{\gamma}_1,\dots,\widetilde{\gamma}_M)$.
The activation function for the each of the non-output layers can be any continuously differentiable non-linear function
within some practical constraints~\cite{B_DEEPLEARNING}.
For the output layer, we choose sigmoid activation function $\bm{\phi}^{(L)}(x) = 1/(1+e^{-x})$
and interpret the network outputs as the predicted FEPs, i.e., $\bm{\eta}^{(L)}\triangleq \bm{\hat{\bm\rho}^\text{NN}}=(\hat{\rho}_1^\text{NN},\dots,\hat{\rho}_K^\text{NN})$.
We choose the cross-entropy loss between the network outputs and the frame error events as the cost function, i.e.,
\begin{align}
\mathcal{C}(\hat{\bm\theta}) &= \frac{1}{K}\sum_{k=1}^K e_k\ln\hat{\rho}_k^\text{NN} + (1-e_k)\ln(1-\hat{\rho}_k^\text{NN}) \\
                                                 &= \frac{1}{K}\sum_{k=1}^K \ln P_{E_k|\bm\Gamma}(e_k|\bm\gamma;\hat{\bm\theta}) \label{eq:nn-cost},
\end{align}
where the latter expression is obtained using Eq.~(\ref{eq:bern-dist}). By observing the direct correspondence between Eqs.~(\ref{eq:cost-func}) and~(\ref{eq:nn-cost}),
we observe that minimizing the neural network cost function over $n=1,\dots,N$ training frames is equivalent to ML estimation of the neural network parameters,
which we have shown to estimate the true FEPs.


In is crucial to point out that there is no guarantee that the neural network trained using stochastic gradient decent will converge to a ML estimates of the parameters,
however, our simulation results indicate that the neural network does indeed provide good estimates of $\rho_k(\bm{\gamma};\hat{\bm{\theta}})$.
Further, the result obtained above
is equivalent to the result derived in the context of cross entropy cost function in~\cite{J_Richard_1991}. However,
we believe that it is instructive to demonstrate this result in the context of ML estimation.


\subsection{FEP Prediction for Throughput Maxmization}
\label{sec:RateSel}

The FEP prediction studied in this paper can be used to select the transmission parameters in each frame
to optimize one or more desired link metrics. In this paper, we study the selection of the optimal channel
code rate $\mathcal{R}_k,k\in \{1,\dots,K\}$  that maximizes the link throughput over a fading channel,
while rest of the transmission parameters are kept fixed. For this, the channel code rate that
maximizes the predicted expected throughput in that frame is selected for each approach, i.e.,
$k^{\text{EESM}} = \arg\max_k T_k(1-\rho_k^\text{EESM})$,
and $^{\text{NN}} = \arg\max_k T_k(1-\rho_k^\text{NN})$
for the EESM and deep neural network approaches respectively, where $T_k=MSJ\mathcal{R}_k$ is the number of trasnmitted information bits when the 
$k^\text{th}$ channel code rate is selected.
The realized throughput over $N$ evaluted frames is therefore $\mathcal{T}^\text{EESM}=\frac{1}{N}\sum_n T_{k^\text{EESM}}^n(1-e_{k^\text{EESM}}^n)$ and 
$\mathcal{T}^\text{NN}=\frac{1}{N}\sum_n T_{k^\text{NN}}^n(1-e_{k^\text{NN}}^n)$
 respectively, where $e_k^n$ denotes the actual frame error event for the $k^\text{th}$ channel code rate in the $n^\text{th}$ frame.


\section{Numerical Results}
\label{sec:Evaluations}

In this section we provide simulation results for the FEP prediction accuracy and
the achieved link throughput for our proposed deep learning approach and contrast it with the EESM approach results.
The simulation parameters are listed in Table~\ref{tab:SimParameters} 
We use Python wrappers over the IT++ library to generate the results described in this section, which allows easy integration
of Tensorflow for the neural network implementation and evaluations~\cite{tensorflow2015-whitepaper}~\cite{PY_ITPP}.
We assume perfect knowledge of the channel at the transmitter as well as the receiver, i.e., $\bm{\gamma}= |\bm{h}|^2/\sigma^2$,
and use a rate-$1/3$ Turbo channel encoder.
The neural network comprises $3$ hidden layers with dimensions $\{60,10,60\}$ respectively and each hidden layer employs a Rectified Linear Unit (ReLU)
activation function~\cite{B_DEEPLEARNING}. The training datasets for EESM and neural network approaches are generated using
$10^4$ frames for each channel code rate.
The test dataset for throughput maximization comprises $10^3$ realizations for $10$ evenly spaced
long term average SINR values in the range $\left[-10,20\right)$ dB.


\begin{table} [!ht]
\centering
\caption{Simulation Parameters}
  \begin{tabular}{|l|c|}
  \hline
  Simulation Parameter  & Value \\
  \hline
  Channel Model         & EPA \\
  Number of Frame OFDM Symbols ($S$) & $12$ \\
  Number of Transmit Subcarriers ($M$) & $600$ \\
  Modulation Order ($J$)           & 2 \\
  Channel Code Rates ($\mathcal{R}_k$)  & $[0.01,0.02...,0.30]$ \\
  \hline
  \end{tabular}
\label{tab:SimParameters}
\end{table}

\begin{figure}[!ht]
\centering
\includegraphics[width=\columnwidth]{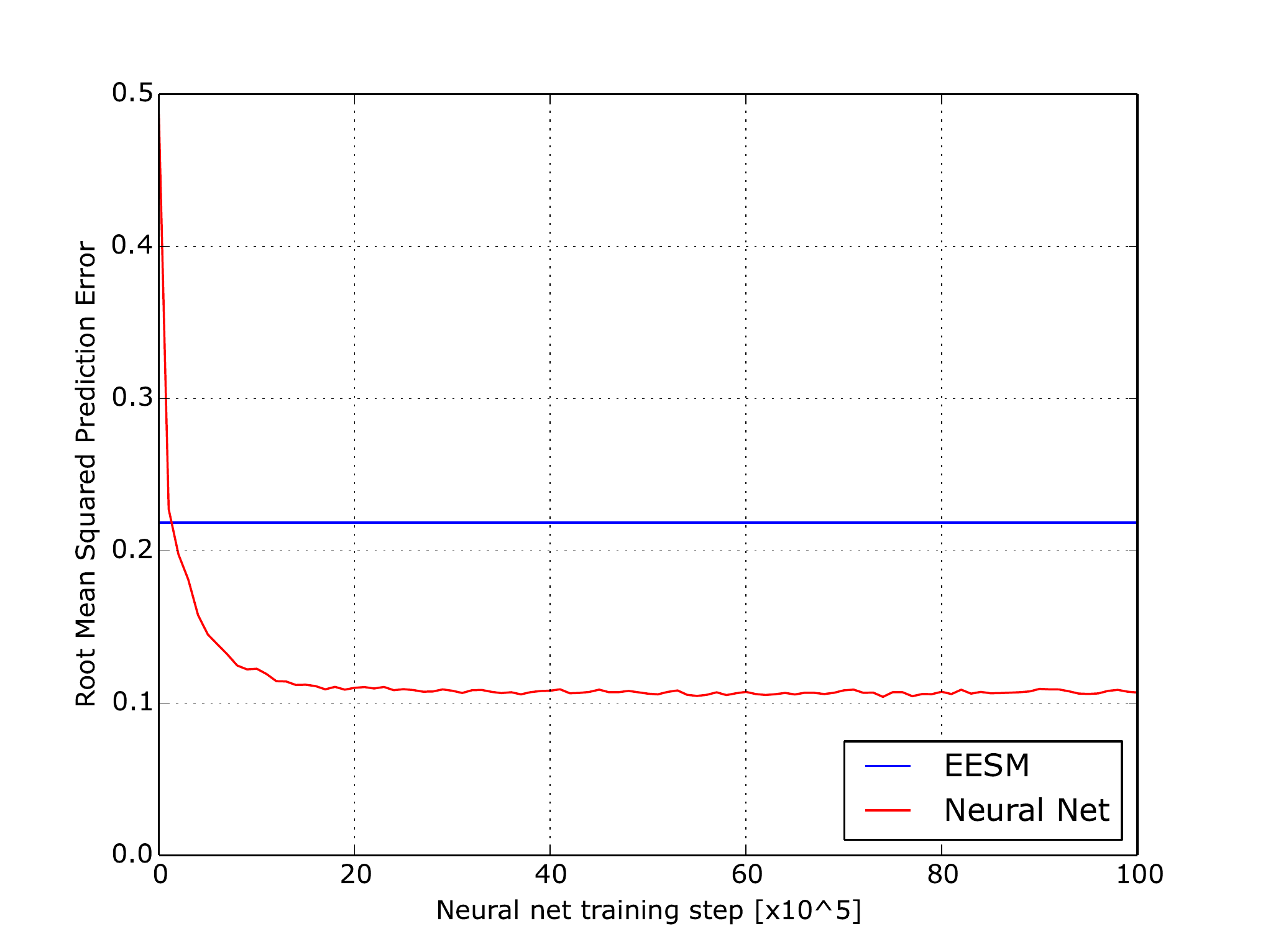}
\caption{RMSE of the FEP prediction error using EESM and Neural Network approaches.}
\label{fig:PRED_MSE}
\end{figure}

\begin{figure}[!ht]
\centering
\includegraphics[width=\columnwidth]{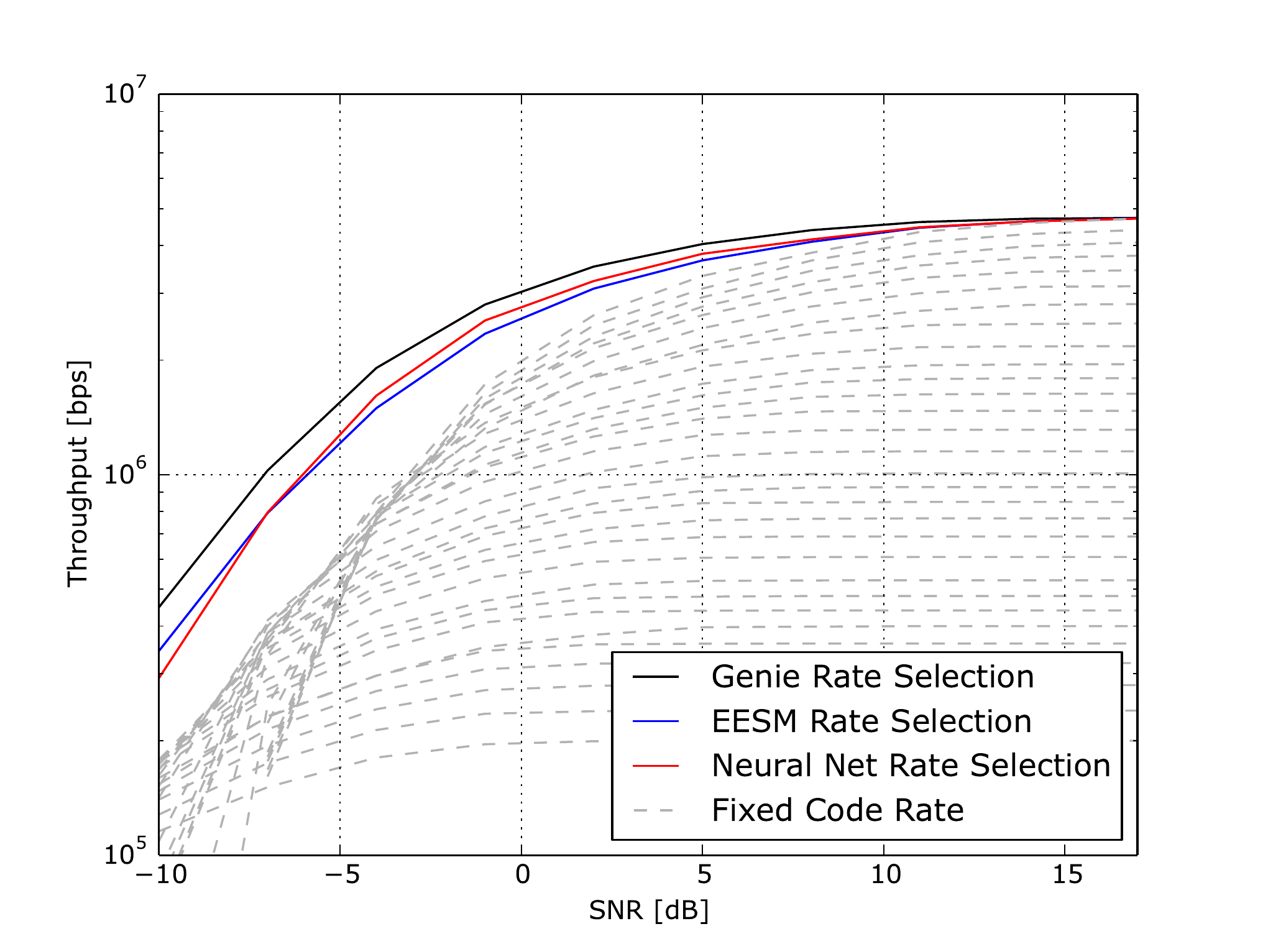}
\caption{Achieved throughput with Genie, EESM and Neural Network rate selection for optimal 
channel code rate selection.}
\label{fig:TPUT_vs_SNR}
\end{figure}

The Root Mean Square Error (RMSE) for FEP prediction performance over is shown in Fig.~\ref{fig:PRED_MSE}.
We observe that the neural network approach outperforms the EESM approach in terms of FEP prediction performance.
The throughput performance of the EESM and neural network approaches
is shown in Fig.~\ref{fig:TPUT_vs_SNR}, along with the upper bound ``Genie'' throughput curve that is obtained
by exhaustively searching the
maximum achieved throughput in each frame. We observe that our approach
increases the throughput compared to the EESM approach.



\section{Conclusions}
\label{sec:Conclusion}

In this paper we have proposed a deep learning approach for FEP prediction 
which directly learns the mapping between the high-dimensional channel state characterized by the per-subcarrier SINRs and the FEPs
for an arbitrary set of transmission configurations. 
Further, by utilizing a training scheme that relies only on the oberved channel state and the binary frame error events,
our approach is able to estimate the true FEPs. 
Finally, we have shown that by exploiting better FEP predictions from our approach, it is possible to increase the realized link throughput by optimally
selecting the channel code rate in each frame.

\vfill\pagebreak



\bibliographystyle{IEEEtran}
\bibliography{refs}

\end{document}